# Nodeless electron pairing in $CsV_3Sb_5$-derived kagome superconductors

Yigui Zhong[1,†], Jinjin Liu[2,8,†], Xianxin Wu[3,†], Zurab Guguchia[4], J.-X. Yin[5,10], Akifumi Mine[1], Yongkai Li[2,7,8], Sahand Najafzadeh[1], Debarchan Das[3], Charles Mielke III[3], Rustem Khasanov[3], Hubertus Luetkens[3], Takeshi Suzuki[1], Kecheng Liu[1], Xinloong Han[6], Takeshi Kondo[1,13], Jiangping Hu[7], Shik Shin[1,11,§], Zhiwei Wang[2,8,9,*], Xun Shi[2,8,*], Yugui Yao[2,8,9], Kozo Okazaki[1,12,13,*]

[1]*Institute for Solid States Physics, The University of Tokyo, Kashiwa, Chiba 277-8581, Japan*

[2]*Centre for Quantum Physics, Key Laboratory of Advanced Optoelectronic Quantum Architecture and Measurement (MOE), School of Physics, Beijing Institute of Technology, Beijing 100081, China*

[3]*CAS Key Laboratory of Theoretical Physics, Institute of Theoretical Physics, Chinese Academy of Sciences, Beijing 100190, China*

[4]*Laboratory for Muon Spin Spectroscopy, Paul Scherrer Institute, CH-5232 Villigen PSI, Switzerland*

[5]*Laboratory for quantum emergence, Department of Physics, Southern University of Science and technology, Shenzhen, Guangdong 518055, China*

[6]*Kavli Institute of Theoretical Sciences, University of Chinese Academy of Sciences, Beijing 100190, China*

[7]*Beijing National Laboratory for Condensed Matter Physics and Institute of Physics, Chinese Academy of Sciences, Beijing 100190, China*

[8]*Beijing Key Lab of Nanophotonics and Ultrafine Optoelectronic Systems, Beijing Institute of Technology, Beijing 100081, China*

[9]*Material Science Center, Yangtze Delta Region Academy of Beijing Institute of Technology, Jiaxing 314011, China*

[10]*Quantum Science Center of Guangdong-Hong Kong-Macao Greater Bay Area (Guangdong), Shenzhen 518045, China*

[11]*Office of University Professor, The University of Tokyo, Kashiwa, Chiba 277-8581, Japan*

[12]*Material Innovation Research Center, The University of Tokyo, Kashiwa, Chiba 277-8561, Japan*

[13]*Trans-scale Quantum Science Institute, The University of Tokyo, Bunkyo, Tokyo 113-0033, Japan*

[†]These authors contributed equally to this work.

[*]Corresponding author. Email: okazaki@issp.u-tokyo.ac.jp; shixun@bit.edu.cn; zhiweiwang@bit.edu.cn

**The newly discovered kagome superconductors represent a promising platform for investigating the interplay between band topology, electronic order, and lattice geometry[1-9]. Despite extensive research efforts on this system, the nature of the superconducting ground state remains elusive[10-17]. In particular, consensus on the electron pairing symmetry has not been achieved so far[18-20], in part owing to the lack of a momentum-resolved measurement of the superconducting gap structure. Here we report the direct observation of a nodeless, nearly isotropic, and orbital-independent superconducting gap in the momentum space of two exemplary $CsV_3Sb_5$-derived kagome superconductors — $Cs(V_{0.93}Nb_{0.07})_3Sb_5$ and $Cs(V_{0.86}Ta_{0.14})_3Sb_5$, using ultrahigh resolution and low-temperature angle-resolved photoemission spectroscopy (ARPES). Remarkably, such a gap structure is robust to the appearance or absence of charge order in the normal state, tuned by isovalent Nb/Ta substitutions of V. Moreover, we observe a signature of the time-reversal symmetry (TRS) breaking inside the superconducting state, which extends the previous observation of TRS-breaking CDW in the kagome lattice. Our comprehensive characterizations of the superconducting state provide indispensable information on the electron pairing of kagome superconductors, and advance our understanding of unconventional superconductivity and intertwined electronic orders.**

Superconductivity often emerges in the vicinity of other ordered electronic states with a broken symmetry, such as antiferromagnetic order and charge density wave. Their interdependence has been widely studied in cuprate and iron-based superconductors[21,22], while persists as a key issue for understanding high-temperature superconductivity. In certain cases, the ordered state and superconductivity can even coexist[23,24], which may indicate an unconventional pairing and have a dramatic impact on the superconducting mechanism. Because of the unique lattice geometry and unusual electronic features in a kagome lattice[1,3,4], the recently discovered kagome superconductors stand out as a new platform for inspecting the superconductivity emerging from a complex landscape of electronic orders[5,6,25,26]. Of particular interest is the nonmagnetic family of $AV_3Sb_5$ ($A$ = K, Rb, Cs)[5,27], in which a variety of intriguing phenomena have been uncovered, including a tantalizing TRS-breaking charge density wave (CDW) order[9,28-30], a pair density wave[10], electronic nematicity[8,31,32], double superconducting domes under pressure[33,34] and giant anomalous Hall effect[35,36]. All these phenomena point out exotic intertwined effects in kagome superconductors $AV_3Sb_5$.

To illuminate the microscopic pairing mechanism and the cooperation/competition between multiple phases in such kagome superconductors, a fundamental issue is to determine the superconducting (SC) gap symmetry. This prominent issue remains elusive owing to the great challenge of resolving such small energy scales and the existence of several conflicting experimental results. Taking $CsV_3Sb_5$ as an example, certain V-shaped gap as well as residual Fermi level states measured by scanning tunnelling spectroscopy[10,11,28] and a finite residual thermal conductivity towards zero temperature[12] seem to support a nodal SC gap. In contrast, the observations of the Hebel-Slichter coherence peak in the spin-lattice relaxation rate from the $^{121/123}$Sb nuclear quadrupole resonance measurement[13] and the exponentially temperature-dependent magnetic penetration depth[14,15], are more consistent with a nodeless superconductivity. On the theoretical side, both unconventional and conventional superconducting pairing were proposed[18-20]. Therefore, an unambiguous characterization of the SC gap structure and its connection with the intertwined CDW order becomes an urgent necessity. During the long-term research of superconductors, ARPES has been proved to be a powerful tool to directly measure the SC gap in the momentum space[37,38]. Nevertheless, the relatively low transition temperature ($T_c$) and correspondingly small gap size render a thorough ARPES measurement extremely challenging.

In this work, we utilize an ultrahigh resolution and low temperature laser-ARPES, together with a chemical substitution of V in $CsV_3Sb_5$ that raises $T_c$, to precisely measure the gap structure in the superconducting state. $CsV_3Sb_5$ crystallizes in a layered structure with V atoms forming a two-dimensional kagome net, as shown in the inset of Fig. 1a. At low temperatures, the material exhibits a CDW transition at $T_{CDW} \sim 93$ K, and eventually becomes superconducting at $T_c \sim 3$ K. To finely tune the competition between superconductivity and CDW, we take two elements to substitute V in $CsV_3Sb_5$. As shown in Fig. 1, both substitutions show a similar trend in the phase diagram, but with distinctions — Nb substitution enhances $T_c$ more efficiently, while Ta dopant concentration can be increased to fully suppress the CDW order. Considering the accessibility in terms of temperature and the possible influence of CDW, we select $Cs(V_{0.93}Nb_{0.07})_3Sb_5$ and $Cs(V_{0.86}Ta_{0.14})_3Sb_5$, from two typical regions in the phase diagram, for the SC gap measurement (denoted hereafter as Nb0.07 and Ta0.14, respectively). The Nb0.07 sample exhibits a $T_c$ of 4.4 K and a $T_{CDW}$ of 58 K, while the Ta0.14 sample exhibits a higher $T_c$ of 5.2 K, but no clear CDW transition. Strikingly, as we shall present below, the gap structures of both samples are isotropic, regardless of the disappearance of CDW, hinting at a robust nodeless pairing in $CsV_3Sb_5$-derived kagome superconductors.

Mapping out the Fermi surface (FS) is critical to investigate the SC gap structure, especially for a multiband system. Due to the limited detectable momentum area of the 5.8-eV laser source, Fig. 2a shows a joint FS of the Ta0.14 sample by combing three segments, which is validated by a whole FS mapping with a higher photon energy (Extended Data Fig. 1). Similar to the pristine $CsV_3Sb_5$ sample[6,39,40], Ta0.14 sample has a circular electron-like pocket (marked as α) and a hexagonal hole-like pocket (marked as β) at the Brillouin zone (BZ) centre Γ point, and a triangle pocket (marked as δ) at the BZ corner K point. The α FS is formed by Sb $5p$ orbitals, while the β and δ FSs are derived from V $3d$ orbitals[39] and are close in momentum. As shown in Figs. 2a and 3b, the β and δ FSs are well distinguished due to the high momentum resolution of the laser source. Moreover, the intensities of β and δ FSs are enhanced under different polarizations of light (supplementary Note 1), which further makes the determination of the Fermi momentum ($k_F$) reliable.

Before investigating the SC gap structure, we first present the spectral evidence of the superconductivity below $T_c$. Using the Ta0.14 sample as an example, the temperature dependent energy distributed curves (EDCs) at $k_F$ of a cut indicated in Fig. 2a are shown in Fig. 2b. At $T = 2$ K far below $T_c$, the emergence of the particle-hole symmetric quasiparticle peaks around Fermi level ($E_F$) clearly indicates the opening of an SC gap. With temperature gradually increasing, the growing intensity at $E_F$ and the approaching quasiparticle peaks suggest that the SC gap becomes smaller and eventually closes. Quantitatively, the SC gap amplitude can be extracted from the fitting procedure based on a Bardeen-Cooper-Schrieffer (BCS) spectral function (supplementary Note 2). The inset of Fig. 2b summarizes the SC gap amplitudes Δ($T$) at different temperatures, which is fitted well with the BCS-like temperature function. The fitted SC gap amplitude at zero temperature, $Δ_0$, is ~ 0.77 meV, and the estimated $T_c$ of ~ 5.2 K is consistent with the bulk $T_c$ determined by resistivity measurement (Fig. 1c). These results demonstrate the high quality of the samples and the high precision of our SC gap measurements.

We then study the momentum dependence of the SC gap in the Ta0.14 sample, in which the CDW order is fully suppressed (Fig. 1b). Considering the six-fold symmetry of the FSs, we select various $k_F$ points to cover the complete FS sheets and thus to capture the symmetry of the SC gap, as shown in Fig. 2f. The EDCs at $k_F$ of the α, β and δ FSs are presented in Figs. 2c–e, respectively. For each $k_F$ point, we take spectra below and above $T_c$, to ensure a precise *in-situ* comparison. In the vicinity of $E_F$, the leading edge of the EDCs at 2 K all show a shift compared to that at 7 K. Moreover, they universally show a strong coherence peak at a binding energy $E_B$ of ~ 1 meV, indicating a rather isotropic SC gap structure. Fitting these EDCs to a BCS spectral function, the quantitatively extracted SC gap amplitudes are summarized in Fig. 2g. These SC gaps of different FSs have rarely fluctuated amplitudes with an average $\Delta_{Ta}$ of 0.77 ± 0.06 meV, yielding a ratio $2\Delta_{Ta}/k_BT_c$ of 3.44 ± 0.27, which is close to the BCS value of ~3.53. These results clearly demonstrate an isotropic SC gap in the Ta0.14 sample.

Next, we turn to examine the possible influence of the CDW order in the normal state on the superconducting pairing symmetry[16,33,34]. We measure the SC gap structure of the Nb0.07 sample, where $T_{CDW}$ gets slightly suppressed, and $T_c$ smoothly increases from that of the pristine CsV$_3$Sb$_5$ (Fig. 1a). In this sense, the superconductivity in the Nb0.07 sample is expected to have a similar SC gap structure with CsV$_3$Sb$_5$. As shown in Fig. 3a, the FS topology of the Nb0.07 sample is also similar to that of CsV$_3$Sb$_5$, consisting of the circular α FS, hexagonal β FS and triangular δ FS, which is consistent with the other ARPES measurements[41] (Extended Data Fig. 1) and the calculations based on density function theory[42]. The EDCs at $k_F$ positions indicated in Fig. 3f, on these three FSs, are presented in Figs. 3c–e, respectively. Just like the case of the Ta0.14 sample, coherence peaks raise up at a similar energy position for all EDCs at 2 K, albeit of a slightly broader shape due to a smaller SC gap and lower $T_c$. By fitting the EDCs to the BCS spectral function, the SC gap amplitudes along the FSs are summarized in Fig. 3g. The data clearly shows a nearly isotropic SC gap structure in the Nb0.07 sample, with the gap amplitude $\Delta_{Nb}$ of 0.54 ± 0.06 meV, giving a ratio $2\Delta_{Nb}/k_BT_c$ of 2.83 ± 0.32, which is smaller than the BCS value. Our results show that an isotropic SC gap robustly persists in the Nb0.07 sample with CDW order.

As the kagome metals $A$V$_3$Sb$_5$ have a three-dimensional electronic structure[39,40], we further study the $k_z$ dependence of the SC gap by tuning the photon energy from 5.8 eV to 7 eV. We find that the SC gap remains nearly the same at these two $k_z$ planes within our experimental uncertainties (Extended Data Fig. 2). Giving the direct momentum-resolving capability of ARPES and the prominent features of SC gap opening, our data reveal a nodeless, nearly isotropic and orbital-independent SC gap in both Nb0.07 and Ta0.14 samples (Figs. 4a, b).

These results shine a light on the interplay between superconductivity and CDW in CsV$_3$Sb$_5$. As shown in Fig. 4c and Extended Data Fig. 3, the isovalent substitutions of Nb/Ta for V in our experiments can be viewed as an effective in-plane negative pressure, which suppresses the CDW order while enhances the superconductivity. In the absence of the CDW order, our measurements on the Ta0.14 sample reveal a nearly isotropic gap structure (Fig. 4b). When the CDW order associated with an anisotropic gap[43,44] comes into play in the Nb0.07 sample, the SC gap remains isotropic and nodeless (Fig. 4a), different from the muon spin relaxation (μSR) observation of a nodal-to-nodeless transition in its sister compounds $KV_3Sb_5$ and $RbV_3Sb_5$ under hydrostatic pressure[16]. The difference between two regimes, represented by Nb0.07 and Ta0.14 samples, is that the ratio $2\Delta/k_BT_c$ is smaller when superconductivity emerges inside the CDW order. This may be attributed to the CDW order partially gapping out the FSs and generating spin polarizations before entering the superconducting phase.

Such robust isotropic SC gaps with small $2\Delta/k_BT_c$ seem to be consistent with a conventional s-wave pairing. However, the coexistence of the TRS-breaking CDW[9,29,30] and superconductivity tends to indicate an unconventional nature. Since TRS is already broken by CDW at $T_{CDW}$ ($\gg T_c$), it is challenging to unambiguously conclude whether the SC state also breaks TRS in $CsV_3Sb_5$ and Nb0.07. To avoid the background from CDW, we performed μSR measurements on the Ta0.14 sample to explore the possible TRS breaking in the pairing state. Intriguingly, as shown in Extended Data Fig. 4, the temperature-dependent zero-field muon spin relaxation rate of the Ta0.14 sample shows a clear enhancement upon the superconducting transition, implying a TRS breaking inside the superconducting state[45]. Thereby, the observed isotropic gap and TRS-breaking nature together suggest a candidate pairing of $s+is$-wave[46] or chiral $d/p$-wave[2,18,19]. Such TRS-breaking superconductivity, naturally connecting with the previously reported TRS-breaking CDW[9,29,30], calls for a concerted mechanism involving multiple interactions (see "Discussion on paring mechanism" in Methods). From a broader perspective, a proper cooperation of multiple interactions might be a common ingredient in promoting unconventional superconductivity[23,24]. Distinguishing from high-$T_c$ superconductivity in the square lattice, such as in cuprate and iron-based superconductors[21,22], the unique lattice geometry of kagome superconductors generates the intrinsic sublattice textures at van Hove fillings, and this promotes nonlocal electronic correlation effect, which could account for the TRS breaking in both superconducting and CDW ordering[2-4]. Our observations taken together provide crucial insights and foundations for further understanding the nature and origin of unconventional superconductivity and will inspire new advances in exploring intertwined electronic orders in kagome quantum materials.

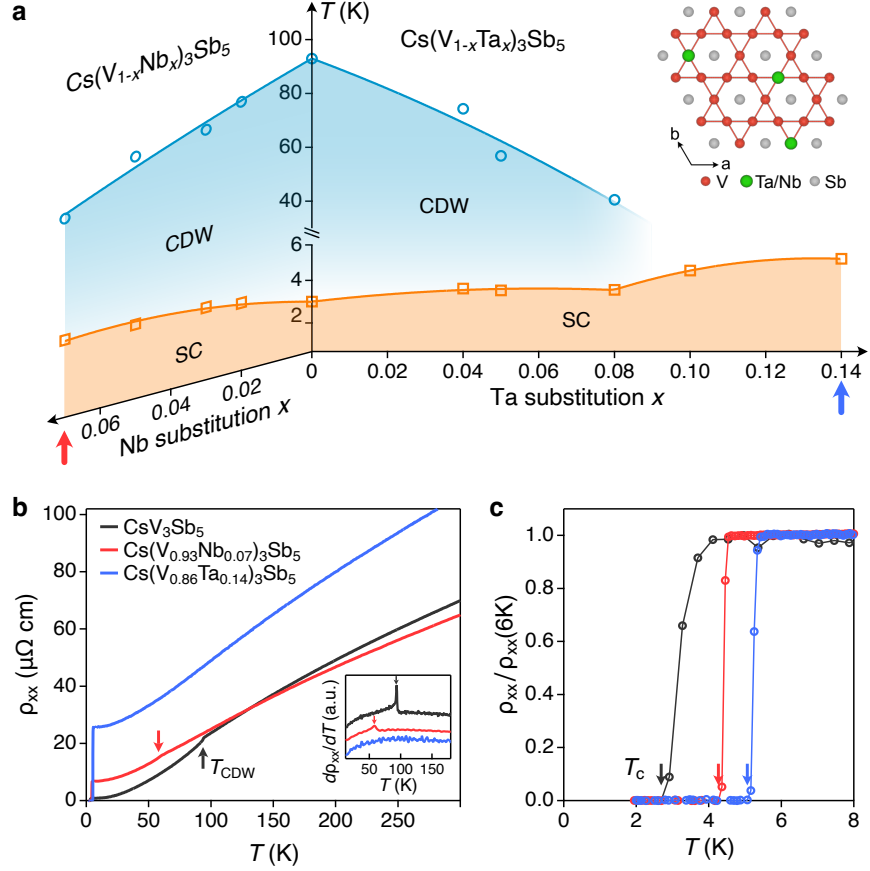

**Fig. 1. Evolution of CDW and superconductivity in CsV$_3$Sb$_5$ upon chemical substitutions. a**, Phase diagrams for Cs(V$_{1-x}$Nb$_x$)$_3$Sb$_5$ and Cs(V$_{1-x}$Ta$_x$)$_3$Sb$_5$. Inset: the lattice structure of V-Sb layer, illustrating the Ta or Nb substitution of V atoms within the kagome lattice. **b**, Temperature dependence of in-plane resistivity for the pristine and two substituted samples studied in this work. The arrows indicate the anomalies associated with CDW transitions. The inset shows the differential resistivity to highlight the CDW transitions, with the curves vertically shifted for clarity. Note that there is no CDW order observed in Cs(V$_{0.86}$Ta$_{0.14}$)$_3$Sb$_5$. **c**, Normalized resistivity curves in the low temperature range showing clear superconducting transitions.

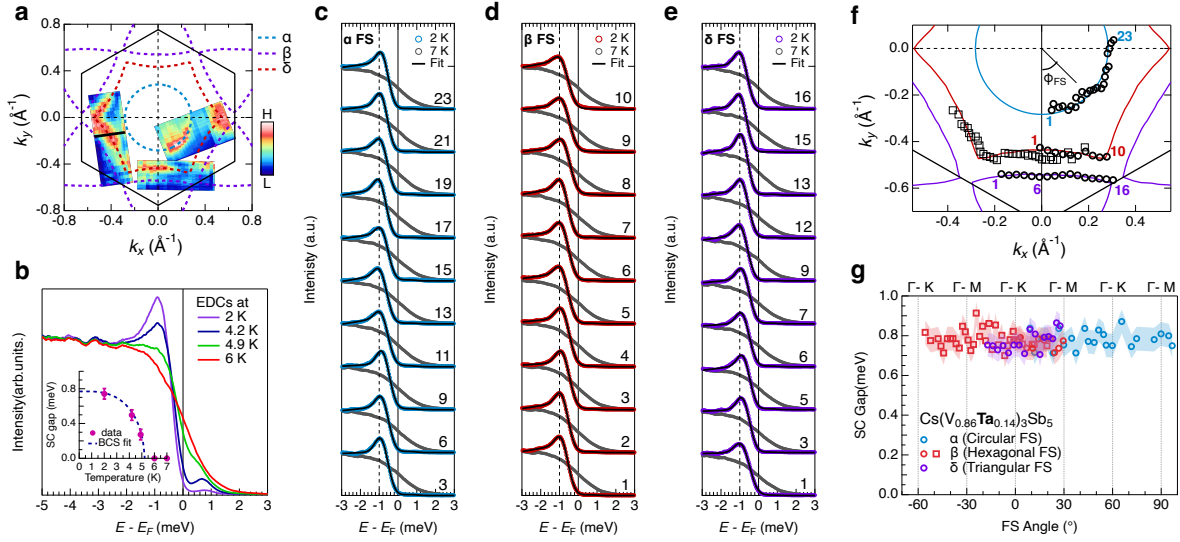

**Fig. 2. Isotropic superconducting gap in Cs(V$_{0.86}$Ta$_{0.14}$)$_3$Sb$_5$. a**, ARPES intensity integrated over ± 5 meV around $E_F$. The broken lines represent the FS contours. **b**, Temperature dependence of EDC at $k_F$ in a cut marked as black line in **a**. Inset shows the temperature dependent SC gap amplitude determined by the fitting procedure based on the BCS spectral function. The blue broken curve represents BCS-like temperature dependence. **c-e**, EDCs at $k_F$ measured at $T$ = 2 K and 7 K along with the α, β and δ FSs, respectively. The $k_F$ positions of these EDCs are summarized in **f** as black thick circles. The black lines are the curves fitted by BCS spectral function. The dashed lines mark the peak of the EDCs. **g**, SC gap magnitude estimated from the fits to EDCs shown in **c-e**. The shaded areas represent the error bars determined from the standard deviation of $E_F$. The square makers are the SC gap results from an independent sample and the corresponding $k_F$ are shown as thin square in **f**.

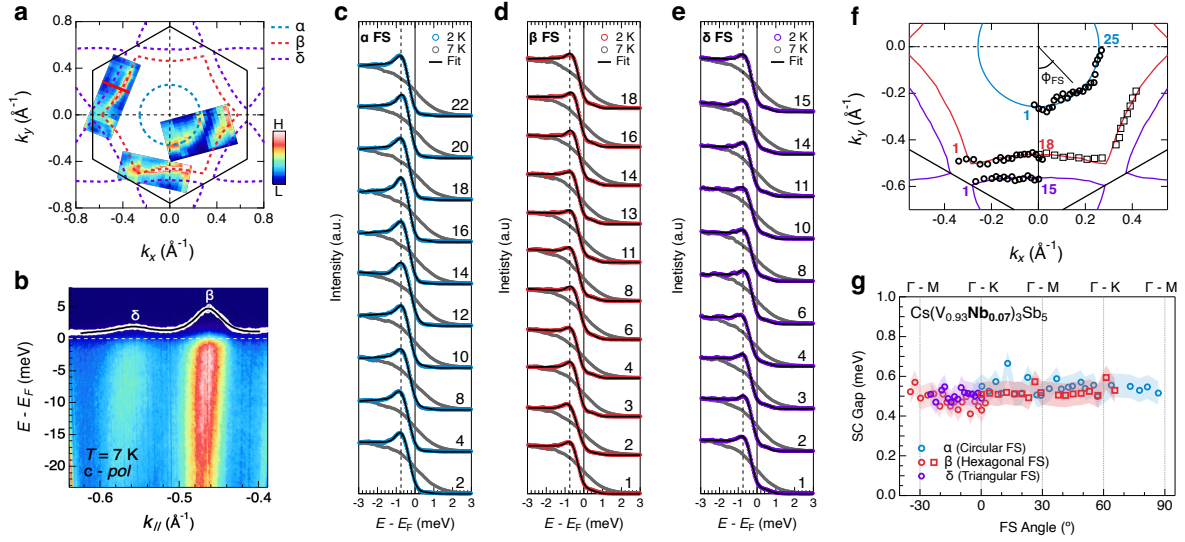

**Fig. 3. Isotropic superconducting gap in charge ordered Cs(V$_{0.93}$Nb$_{0.07}$)$_3$Sb$_5$. a**, ARPES intensity integrated over ± 5 meV around $E_F$. The broken lines represent the FS contours. **b**, ARPES intensity plot along a red line shown in **a**. The intensity is measured using circular polarization to capture both β and δ bands. The white dotted line represents the MDC integrated over ± 2 meV around $E_F$ and the black line is a double-peak Lorentzian fit. Two distinguished peaks in the MDC shows $k_F$ positions of the β and δ bands. **c-e**, EDCs at $k_F$ taken along the α, β and δ FSs, respectively. The $k_F$ positions of these EDCs are summarized in **f**. The dashed lines mark the estimated peak position of the EDCs. The black lines are the curves fitted by BCS spectral function. **g**, SC gap magnitude estimated from the fits to EDCs shown in **c-e**. The square makers are the SC gap results from an independent sample and the corresponding $k_F$ are shown as thin square in **f**. The shaded areas represent the error bars determined from the standard deviation of $E_F$.

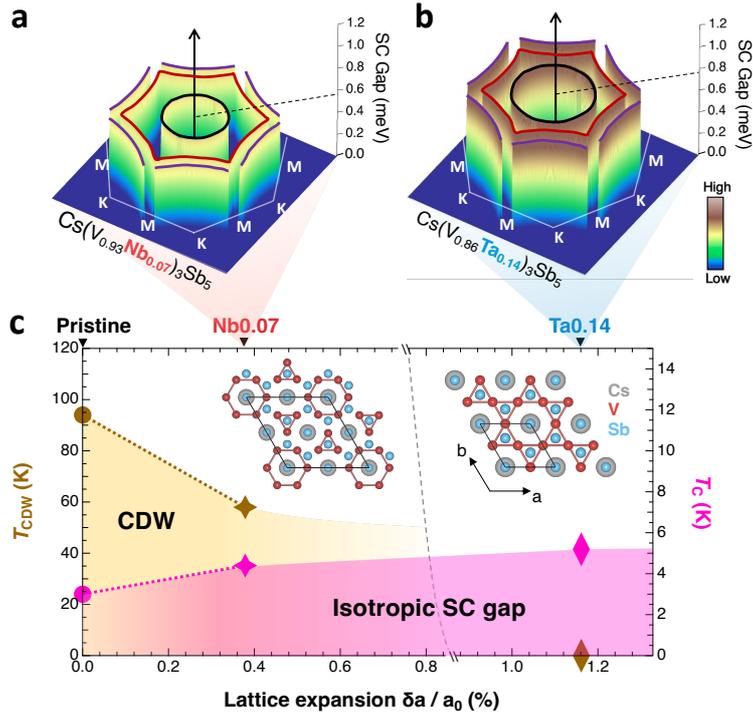

**Fig. 4. Robust isotropic SC gap upon suppression of CDW. a-b,** Schematic momentum dependence of the SC gap magnitude of the Nb0.07 and Ta0.14 samples, respectively. **c,** Schematic phase diagram in which $T_{CDW}$ and $T_c$ are plotted as function of the lattice expansion due to the chemical substitutions. The lattice expansion is represented by $\delta a/a_0$, where $\delta a = a - a_0$ is the change of the in-plane lattice constant $a$ from that of pristine $CsV_3Sb_5$ ($a_0$). The inset shows the lattice structures of the CDW (left) and undistorted (right) phases, representing the states above $T_c$ for two distinct regions in the phase diagram. The black solid lines in the insets mark the corresponding single-unit cells. The isotropic SC gap symmetry persists through such two regions, regardless of the existence of a CDW order.

**Methods**

**Growth of single crystals.** High-quality single crystals of $Cs(V_{0.86}Ta_{0.14})_3Sb_5$ and $Cs(V_{0.93}Nb_{0.07})_3Sb_5$ were synthesized from Cs bulk (Alfa Aesar, 99.8%), V piece (Aladdin, 99.97%), Ta powder (Alfa Aesar, 99.99%), and Sb shot (Alfa Aesar, 99.9999%), via a self-flux method using $Cs_{0.4}Sb_{0.6}$ as flux. The above starting materials were put into an aluminium crucible and sealed in a quartz tube, which was then heated to 1000°C in 24h and dwelt for 200 h. After that, the tube was cooled to 200°C at a rate of 3°C/h, followed by cooling down to room temperature with the furnace switched off. In order to remove the flux, the obtained samples were soaked in deionized water. Finally, shiny single crystals with hexagonal feature were obtained.

**Electronic transport measurements.** Electronic transport properties of $Cs(V_{0.86}Ta_{0.14})_3Sb_5$ and $Cs(V_{0.93}Nb_{0.07})_3Sb_5$ crystals were measured on a physical property measurement system (PPMS, Quantum Design) at a temperature range from 300 K to 1.8 K. Five-terminal method was used, at which the longitudinal resistivity and Hall resistivity can be taken simultaneously. DC magnetic susceptibility was measured on a magnetic property measurement system (MPMS, Quantum Design) with a superconducting quantum interference device (SQUID) magnetometer.

**High-resolution laser-ARPES measurements.** Ultrahigh-resolution ARPES measurements were performed in a laser-based ARPES setup at the ISSP, University of Tokyo, which consisted of a continuous wave laser (h$\upsilon$ = 5.8 eV) provided from OXIDE Corporation and a vacuum ultraviolet laser (h$\upsilon$ = 6.994 eV), a Scienta HR8000 hemispherical analyser, and a sample manipulator cooled by decompression-evaporative the liquid helium. The samples were *in-situ* cleaved and measured under a vacuum better than $3\times10^{-11}$ torr. The sample temperature was varied from 2 to 7 K, and the energy resolution for the superconducting gap measurements was better than 0.6 meV for 5.8-eV laser and 1.5 meV for 6.994-eV laser. We checked the linearity of the detector[47], and the Fermi level $E_F$ was calibrated with an *in-situ* connected gold reference.

**µSR measurements.** Zero field (ZF) µSR experiments were performed on the general-purpose surface-muon (GPS) instrument. When performing experiments in zero-field mode, the compensation is done dynamically to ensure true zero-field conditions independently of the status of the external magnetic field sources. Zero field data analysis were performed using the so-called single-histogram models[16,29,48]. In order to differentiate between static and fluctuating internal magnetic fields, so-called longitudinal-field experiments are also performed where the externally applied magnetic field is along the initial muon spin direction. A mosaic of several crystals stacked on top of each other was used for these measurements.

**Discussion on pairing mechanism.** Our comprehensive results from the combined high-resolution ARPES and μSR measurements provide crucial implications for the pairing mechanism in $CsV_3Sb_5$ kagome family. Due to the unique lattice geometry, the van Hove singularities possess sublattice textures on the Fermi surface, and this will promote the nonlocal electronic correlation effect through the sublattice interference[2-4,18]. In these kagome materials with multiple van Hove singularities associated with V 3$d$ orbitals[39,40], the correlation effect is considered to be important for the appearance of intriguing phenomena. For superconductivity, it will give rise to a nodal or nodeless pairing with strong anisotropy or a chiral $d/p$-wave pairing with an isotropic gap[2,18,19]. In particular, the chiral $d/p$-wave pairing is fully gapped and breaks TRS, consistent with our μSR measurement. In the scenario of the pure electronic interaction, however, the SC gap is expected to exhibit noticeable orbital dependence due to the distinct electronic correlations, which is partially inconsistent with our results. On the other hand, phonon hardening across the CDW transition[49] and the observed band dispersion kinks[43,50] suggest that electron-phonon coupling is non-negligible in these kagome metals. Notably, the determined electron-phonon coupling strength of certain phonon modes is positively correlated with $T_c$ in our measurements (Extended Data Fig. 5). Electron-phonon coupling alone usually generates a uniform $s$-wave pairing, which, however, cannot account for the observed TRS breaking in the superconducting state. In addition, the observed isotropic SC gap with TRS breaking seems to also be consistent with the $s+is$-wave pairing proposed in multi-orbital iron-based superconductors, which is derived from the inter-band scattering driven by the electronic interaction[46]. Therefore, both electronic interaction and electron-phonon coupling could possibly play crucial roles in promoting superconductivity in these kagome superconductors. They may reinforce each other in the frustrated kagome lattice, similar to that in cuprate and iron-based superconductors[23,24], collaboratively generating a TRS broken pairing with an isotropic gap. We also note that when superconductivity coexists with the CDW in the Nb0.07 sample, the pairing is nodeless determined by ARPES measurements, but μSR measurements cannot unambiguously identify whether this pairing breaks TRS or not, which deserves further experimental study.

**Competing interests:** The authors declare no competing interests.

**Data availability:** Data are available from the corresponding author upon reasonable request.

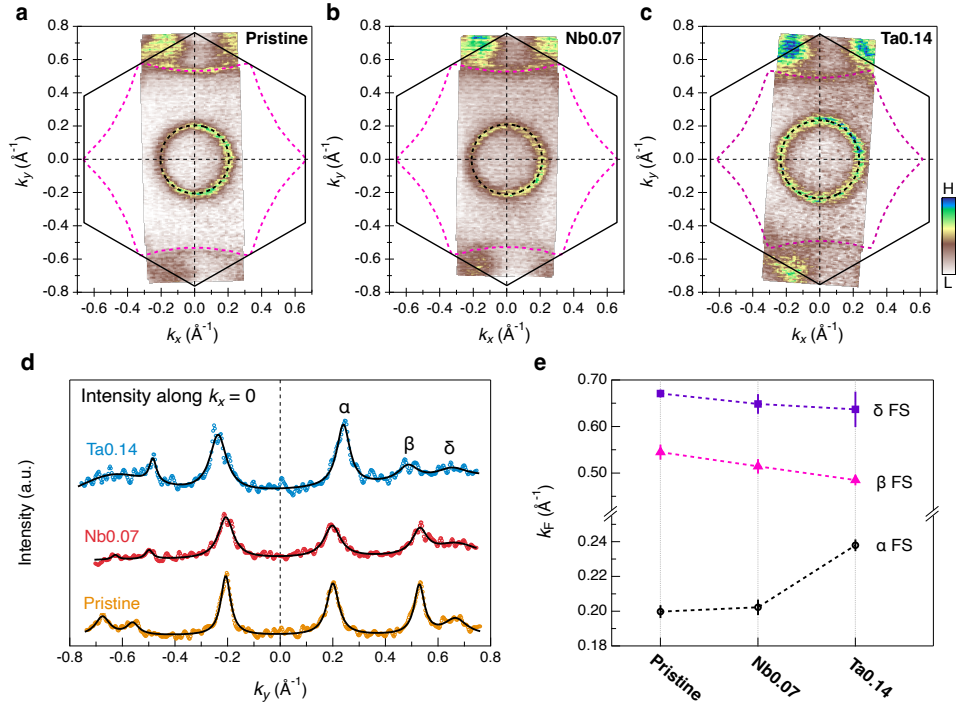

**Extended Data Fig. 1. Fermi surface evolution upon Nb/V substitutions. a-c**, Fermi surface maps integrated over $E_F \pm 5$ meV for the pristine, 7%-Nb and 14%-Ta substituted $CsV_3Sb_5$ samples, respectively. The spectra were measured with He I$\alpha$ photons (hv = 21.218 eV) at $T$ = 7 K. **d**, Line cuts along $k_x = 0$. The black lines are the Lorentizen fits to determine the $k_F$ positions. **e**, Summary of the $k_F$ evolution of three Fermi surfaces upon Nb/V substitutions. The error bars represent the uncertainies of the fits.

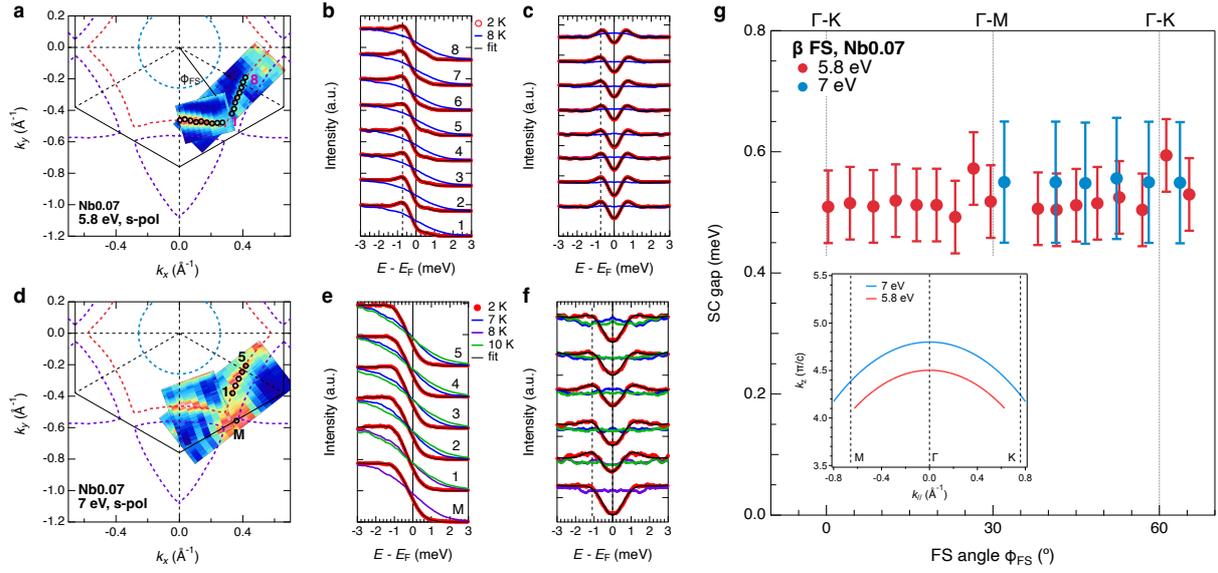

**Extended Data Fig. 2. Superconducting gap at different $k_z$ for the Cs(V$_{0.93}$Nb$_{0.07}$)$_3$Sb$_5$ sample. a**, FS map taken with 5.8-eV laser. **b**, EDCs at $k_F$ marked in **a**. The black lines are the fits of these EDCs. **c**, Symmetrized EDCs for **b**. **d-f**, Same as a-c but for the data taken with 7-eV laser. The curves are vertically offset for clarity. **g**. Comparison of the SC gap amplitude measured with 5.8-eV and 7-eV laser. The inset shows the $k_z$ positions corresponding to these two photon energies.

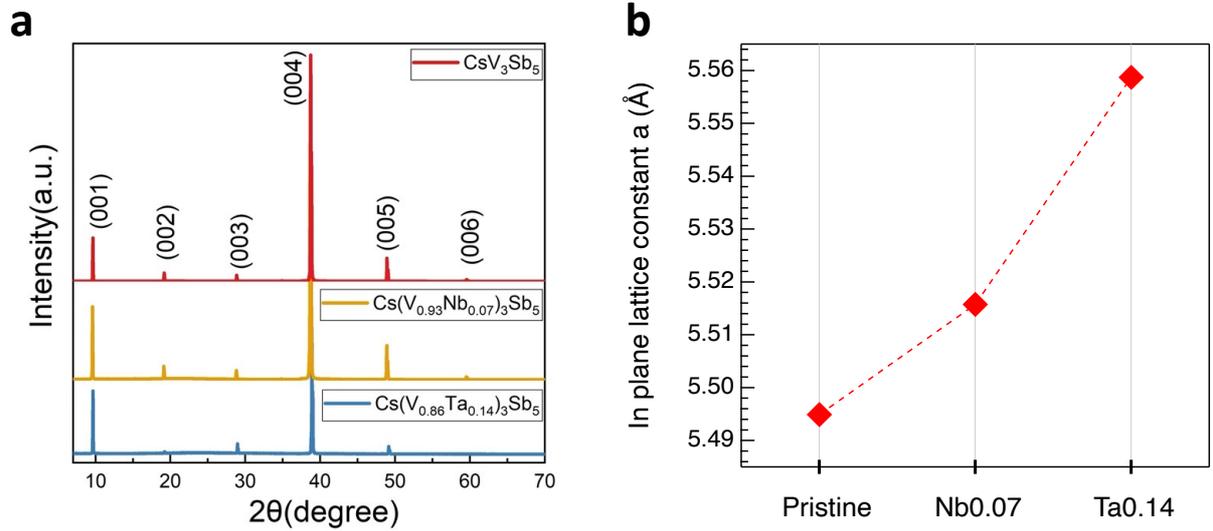

**Extended Data Fig. 3. X-ray diffraction pattern (a) and in-plane lattice constants (b) of the pristine CsV$_3$Sb$_5$, Nb0.07 and Ta0.14 single crystals.**

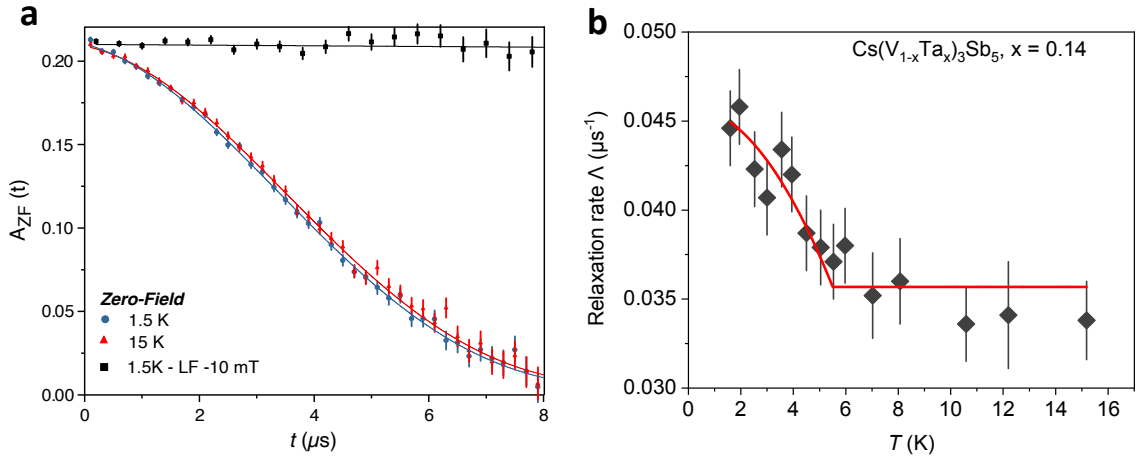

**Extended Data Fig. 4. Time-reversal symmetry breaking in the superconducting state of the Cs(V$_{0.14}$Ta$_{0.86}$)$_3$Sb$_5$ sample with CDW fully suppressed. a,** Zero-Field (ZF) μSR time spectra for Cs(V$_{0.86}$Ta$_{0.14}$)$_3$Sb$_5$ below and above $T_c$. The solid lines are the fits to the data using the Gaussian Kubo–Toyabe depolarization function, which reflects the field distribution at the muon site created by the nuclear moments of the sample, multiplied by an additional exponential $exp(-\Lambda t)$ term[9] (see supplementary note 7). Exponential term accounts for any additional relaxation channels (such as broken TRS). The μSR time spectra (marked as black square) in a longitudinal filed of 10 mT below $T_c$ is plotted to show full decoupling of the muon spin. **b,** Temperature dependence of the zero-field muon spin relaxation rate, $\Lambda$, in the temperature range across $T_c$. The error bars represent the standard deviation of the fit parameters. The red line is a fit with the empirical relation (see supplementary note 7). The relaxation rate increases smoothly below $T_c$, implying the TRS breaking inside the superconducting state.

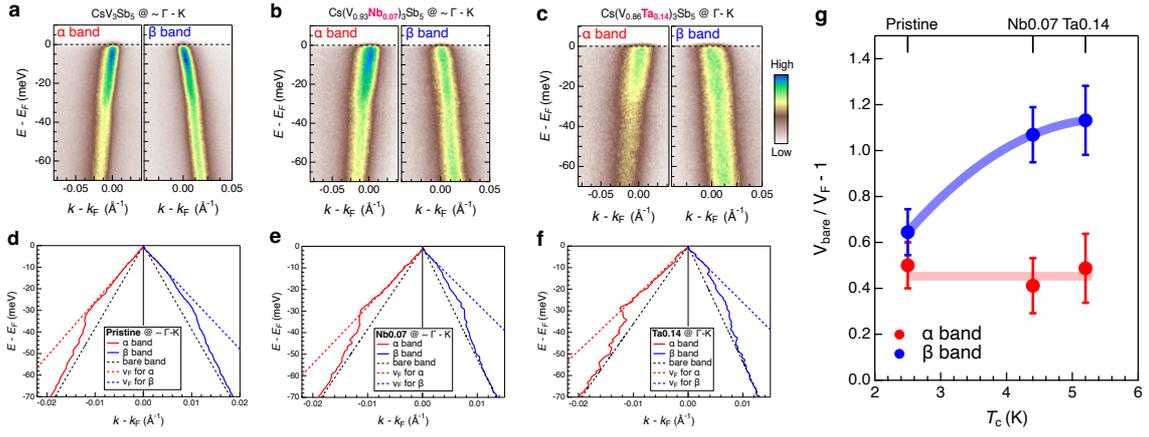

**Extended Data Fig. 5. Spectral evidence of the electron-phonon coupling. a-c**, ARPES intensity plots of the α and β bands nearly along Γ−K direction for the pristine $CsV_3Sb_5$, Nb0.07 and Ta0.14 samples, respectively. These ARPES data are taken with 7-eV laser at $T = 6$ K. **d-f**, Extracted band dispersions. **a** and **d** are adopted from the Reference[50], in which the $T_c$ of the measured $CsV_3Sb_5$ is ~ 2.5 K. **g**, Ratio between the velocity of the bare band and the Fermi velocity for the pristine, Nb0.07 and Ta0.14 samples, plotted as a function of their $T_c$.